\documentclass[12pt,hyper]{JHEP3}
\usepackage{amssymb}
\usepackage{mathrsfs} 
\usepackage{wasysym}
\newcommand{\Barcelo}{Barcel\'o}
\newcommand{\scri}{\mathscr{I}}

\title{Hawking-like radiation from evolving black holes and compact horizonless objects}
\author{
Carlos \Barcelo\\
Instituto de Astrof\'{\i}sica de Andaluc\'{\i}a, IAA--CSIC,
Glorieta de Astronom\'i{}a, \\
18008 Granada, Spain\\
E-mail: {\email{ carlos@iaa.es}}
}
\author{
Stefano Liberati\\
SISSA/ International School for Advanced Studies, Via Bonomea 265,\\
34136 Trieste, Italy \emph{and} INFN, Sezione di Trieste\\
E-mail: \email{liberati@sissa.it}
}
\author{
Sebastiano Sonego\\
Universit\`a di Udine, Dipartimento di Fisica, Via delle Scienze 208, \\
33100 Udine, Italy\\
E-mail: \email{sebastiano.sonego@uniud.it}
}
\author{
Matt Visser\\
School of Mathematics, Statistics, and Operations Research,\\
Victoria University of Wellington, New Zealand\\
E-mail: \email{matt.visser@msor.vuw.ac.nz}

\vskip 5 cm

\thispagestyle{empty}
}

\abstract{

Usually, Hawking radiation is derived assuming (i) that a future eternal event horizon forms, and (ii) that the subsequent exterior geometry is static.  However, one may be interested in either considering quasi-black holes  (objects in an ever-lasting state of approach to horizon formation, but never quite forming one), where (i) fails, or, following the evolution of a black hole during evaporation, where (ii) fails. 
We shall verify that as long as one has an approximately exponential relation between the affine parameters on the null generators of past and future null infinity, then subject to a suitable adiabatic condition being satisfied, a Planck-distributed flux of Hawking-like radiation will occur.  This happens both for the case of an evaporating black hole, as well as for the more dramatic case of a collapsing object for which no horizon has yet formed (or even will ever form).  In this article we shall cast the previous statement in a more precise and quantitative form, and subsequently provide several explicit calculations to show how the time-dependent Bogoliubov coefficients can be calculated.
}

\begin{document}

\def\implies{\Rightarrow}
\renewcommand{\sun}{\ensuremath{\odot}}%
\def\ie{{\emph{i.e.}}}
\def\etc{\emph{etc}}
\section{Introduction}

Ever since Hawking's original 1974 derivation that black holes emit a steady Planck-distributed flux of quanta~\cite{hawking1, hawking2}, there has been a steady and continual stream of articles that re-derive the Hawking flux in various different ways~\cite{unruh, tipler, israel, bardeen, york, hajicek, grove, entropy, ergo, essential, lindesay+sheldon, Parikh:1999, Parikh:2004a, Parikh:2004b, Shankaranarayanan:2000, Angheben:2005, Medved:2005, Arzano:2005, Robinson:2005, Clifton:2008, Banerjee:2009, Padmanabhan:2003, Hossenfelder:2002}  --- the justification being that in doing so one might strip the calculation to its essence and so discover what aspects of black hole physics are truly important for the phenomenon, and what aspects can safely be put aside.  Several important points are by now well established, though often not well appreciated:
\begin{itemize}
\item 
Hawking radiation is more ubiquitous than Bekenstein entropy, and Hawking radiation will still occur in situations where the notion of black hole entropy has no meaning~\cite{entropy,ergo, essential}. 
\item 
Eternal black holes, and their associated bifurcate Killing horizons, are useful mathematical models~\cite{unruh}, and the source for many useful heuristics~\cite{dbh}, but physically they do not accurately reflect the formation and evolution of real astrophysical black holes.
\item 
The subtle differences between various forms of horizon present in general relativity (event, apparent, isolated, dynamical, trapping, \etc...) are physically important~\cite{hajicek, essential, bergmann-roman, quasiparticle, trapping, fate, small-dark, revisit} --- and precisely which (if any) of these is essential for Hawking radiation has direct impact, for instance on the question of ``information loss''~\cite{hawking-dublin, hawking-post-dublin, ashtekar-bojowald, disinformation,  hayward}. 
\end{itemize}
In the current article we will address two specific questions (throughout this paper we will adhere to strictly general relativistic analyses, not considering modified dispersion relations):
\begin{enumerate}
\item[(i)] How can we derive the Hawking flux emitted by a slowly evolving (as opposed to static) black hole-like object? 
\item[(ii)] What type of horizon (if any) is required to generate a Hawking-like Planck-distributed flux of quanta?
\end{enumerate}
We shall do so by adapting, modifying, and extending the by now reasonably well established result that if (in spherical symmetry) you have an (approximately) exponential
relation between the affine parameters $U$ and $u$ on the null generators of past and future null infinity, $\scri^-$ and $\scri^+$, then Hawking
radiation will occur. 
The standard form of this argument (as per Hawking's original calculation~\cite{hawking1, hawking2}) is to say that if asymptotically
\begin{equation}
U \approx U_H - A \; e^{- \kappa_H \, u},
\end{equation}
as $u\to+\infty$  for some arbitrary positive constants $A$ and $\kappa_H$,
then Hawking radiation happens, with fixed non-evolving temperature
\begin{equation}
k_B \;T_H = {\hbar\; \kappa_H\over2\pi}.
\end{equation}
Here $U$ and $u$ can be viewed as different labels that we attach to a null curve connecting $\scri^-$ with $\scri^+$.  We can equivalently write
\begin{equation}
u \approx - {1\over\kappa_H} \; \ln\left({U_H - U\over A}\right).
\end{equation}
Of course the original Hawking calculation~\cite{hawking1, hawking2}  explicitly assumes the existence of
an event horizon at $U_H$, and that the resulting black hole quickly settles down to a static configuration.

In counterpoint, in 1987 Hajicek~\cite{hajicek} demonstrated that a strict \emph{event} horizon was not necessary, and that a (suitably long-lived) \emph{apparent} horizon was quite sufficient to generate a Hawking flux. (See also~\cite{essential, hawking-dublin, hawking-post-dublin}.)   More recently (within the context of ``analogue spacetimes''~\cite{entropy, ergo, unexpected, LRR}) the present authors have demonstrated that apparent/ trapping horizons can also be dispensed with, or their appearance postponed indefinitely into the future~\cite{quasiparticle, trapping}.  
We shall now make these results more general, precise, and quantitative~\cite{our-prl-article}. Inspired in particular by the work of Hu~\cite{hu}, we focus on the existence of an (in our case, approximate) exponential relation between the affine parameters on past and future null infinities as the necessary and sufficient condition for generating a Hawking flux.

Here is a summary of our key result: Consider null curves starting
from $\scri^-$  and arriving on $\scri^+$.  There will be
\emph{some} relation between the affine parameters $U$ on $\scri^-$ and $u$ 
on $\scri^+$:
\begin{equation}
U = p(u);  \qquad u = p^{-1}(U).
\end{equation}
Now pick a particular null curve, labelled by  $U_*$  on $\scri^-$ and 
$u_*$ on $\scri^+$. We can without loss of generality write
\begin{equation}
U = U_* 
+ C_* \;  \int^u_{u^*}  
\exp\left[  - \int^{\bar u}_{u_*} \kappa(\tilde u) \; d\tilde u \right] d \bar u,
\end{equation}
for some constant $C_*$ and the function $\kappa(u) = - \ddot p(u)/\dot p(u)$.
Assume (and this is where the physics comes in) that $\kappa(u)$ satisfies an ``adiabatic condition''
\begin{equation}
\left|\dot \kappa(u_*)\right|  \ll \kappa(u_*)^2.
\label{E:adiabatic0}
\end{equation}
Then we shall show that this is sufficient  to guarantee (under mild technical assumptions) the existence of a Hawking-like Planck-spectrum of
outgoing particles reaching $\scri^+$ at $u_*$, now with a time-dependent
Hawking temperature
\begin{equation}
k_B \; T_H( u_* ) = {\hbar \; \kappa( u_* )\over2\pi}.
\end{equation}
As we move along $\scri^+$ (that is, as $u_*$ increases, possibly even
with $u_* \to +\infty$), this relation continues to hold, with the
Hawking temperature controlled by $\kappa(u_*)$,  \emph{as long as the
adiabatic condition continues to hold}.
We do \emph{not} need to assume a horizon --- \emph{of any sort} --- ever forms.
The rest of this article will be devoted to a detailed proof of this result. 
Along the way we shall revisit (and hopefully clarify) the salient features that go into calculating the relevant Bogoliubov coefficients --- specifically we shall very carefully look at the issue of defining and accurately estimating appropriate \emph{time-dependent} Bogoliubov coefficients.

\section{Motivating the adiabatic condition}

Why might we even expect something like the ``adiabatic condition'' (\ref{E:adiabatic0}) to either be true or relevant? 
We can physically interpret  the ``adiabaticity condition'' as
equivalent to the statement that a photon emitted near the peak of the
Planck spectrum, with $\hbar \, \omega_\infty\approx k_B \, T_H$, that is
$\omega_\infty \sim \kappa$, should \emph{not} see a large fractional
change in the peak energy of the spectrum over one oscillation of the
electromagnetic field. (That is, the change in spacetime geometry is adiabatically slow as
seen by a photon near the peak of the Hawking spectrum.) It is this slow change in the spacetime geometry that ultimately permits us to apply a variant of Hawking's original calculation. 

To then verify that this adiabatic condition holds for macroscopic black holes let us (for example)  think of the standard Hawking
calculation for a Schwarzschild black hole. The Hawking temperature
is~\cite{hawking1, hawking2}:
\begin{equation}
k_B \; T_H \sim \hbar\;  \kappa \sim {M_\mathrm{Planck}^2\over M},
\end{equation}
and consequently (\emph{assuming} self-consistent back-reaction and that the Hawking formula continues to be true for a slowly evolving almost-Schwarzschild black hole) the mass loss rate for a black hole evaporating into vacuum is given by the standard result
\begin{equation}
|\dot M| 
\sim
{M_\mathrm{Planck}^4  \over M^2}.
\end{equation}
In particular, as the black hole evolves its temperature changes. This brings up and reinforces an important point: Any truly fundamental derivation of the Hawking effect should be able to deal with a time-dependent Hawking temperature.  If your favourite derivation is intrinsically incapable of dealing with time dependent situations, then such a derivation is missing fundamental parts of the physics. 

Looking at the surface gravity of the Schwarzschild black hole we can estimate
\begin{equation}
{\dot\kappa\over \kappa^2} \sim
{M_\mathrm{Planck}^2\over M^2}.
\end{equation}
So the standard Hawking process for standard Schwarzschild black holes
does satisfy the ``adiabaticity condition'' we have enunciated above,
\emph{at least as long as the black hole is heavier than a few Planck
masses}.

Overall this now provides us with a coherent physical picture all the way down to the Planck mass, where we see that adiabaticity breaks down, and ``quantum gravity'' (in the sense of ``that quantum theory that approximately reduces to general relativity in an appropriate limit") takes over.

\section{The exponential approximation}
\subsection{Definitions and exact results}

Consider an asymptotically flat spherically symmetric spacetime with a Minkowskian structure in the asymptotic past. (The discussion that follows applies equally well to any number of spatial dimensions and can easily be generalized to deal with acoustic spacetimes in 1+1 dimensions having two asymptotic regions~\cite{njp}.) In the $\{t,r\}$ sector of the geometry we  
define an affine parameter $W$ on $\scri^-$, and use it to label the null curves travelling towards the centre of the body. Similarly, $u$ is taken to be an affine parameter on $\scri^+$, used to label the null rays  travelling away from the central body.  The independent coordinates $\{W,u\}$ provide a double-null cover of the relevant parts of spacetime (the domain of outer communication).

As is standard, one can define a canonical functional relationship connecting $\scri^-$ with $\scri^+$ by using null curves that reflect off the centre at $r=0$. This relation can be expressed as
\begin{equation}
U = p(u),  \qquad u = p^{-1}(U), 
\end{equation}
where the labels $\{U,u\}$ are now no longer to be thought of as independent coordinates but, since we have explicitly linked them via the function $p(\cdot)$, as different ways of labelling the same null curve once it is reflected through the origin.
It is to be understood that $p^{-1}(\cdot)$
need not be defined on all of $\scri^-$ if a true event horizon indeed forms; however this function will certainly be well defined on those parts of $\scri^-$ that lie in the domain of outer communication.
We shall soon see that the function $p(\cdot)$, or equivalently its inverse,
is sufficient to encode all the relevant physics of Hawking radiation. Specifically, let us choose a reference null curve completely traversing the body. It is labelled by $u_*$ on its way out of the body, and by $U_*$ on its way in.  We want to use ``local'' information from the vicinity of this reference null curve to study Hawking-like radiation that reaches $\scri^+$ in the vicinity of $u_*$.

Let us now start the technical computation by \emph{defining} a quantity $\kappa(u)$ via the relation
\begin{equation}
\kappa(u) =  -{d^2 U/du^2\over {dU/du} } = -{\ddot p(u)\over \dot p(u)},
\end{equation}
so that $\kappa(u)$ is simply a functional parameterization of the relationship between the affine parameters $U$ and $u$.  When this function happens to be almost constant it controls the $e$-folding relationship between $u$ and $U$, and so provides a notion of ``surface gravity'' in terms of the ``peeling'' properties of null geodesics. (And so is intimately related  to $\kappa_\mathrm{effective}$ as defined in~\cite{quasiparticle, trapping}. See also~\cite{Macher, Brout1, Brout2} for comments on the importance of these ``peeling'' properties.)
Now pick some generic null curve labelled by $u_*$, 
then through integration one can express any $U= p(u)$ as a function of its corresponding $\kappa(u)$:
\begin{equation}
U = U_* 
+ C_* \;  \int^u_{u^*}  
\exp\left[  - \int^{\bar u}_{u_*} \kappa(\tilde u) \; d\tilde u \right] d \bar u,
\end{equation}
for some constant $C_*$.  Note that it is impossible, even in
principle, to use \emph{local} physics to specify a unique normalization for $C_*$. This is
 ultimately due to the fact that any constant multiple of an affine
null parameter is still an affine null parameter. Note also that $C_*$ is a constant in the sense that it is the same for all null curves $u$ in the vicinity of the null curve specified by $u_*$. However $C_*$ does depend on the choice of the specific null curve $u_*$ one is working around. This formalism continues to make perfectly good sense even for $\kappa=0$, where it simply implies a linear relation between $u$ and $U$.
Let us now write
\begin{equation}
\kappa(u) = \kappa(u_*) + \delta\kappa(u) = \kappa_* + \delta\kappa(u);
\end{equation}
then in particular we have
\begin{equation}
\int^{\bar u}_{u_*} \kappa(\tilde u) \; d\tilde u=   \kappa_* (\bar u - u_*) + \int^{\bar u}_{u_*} \delta\kappa(\tilde u) \; d\tilde u,
\end{equation}
so that (still an exact result)
\begin{equation}
\int^u_{u^*}   \exp\left[  - \int^{\bar u}_{u_*} \kappa(\tilde u) \; d\tilde u \right] d \bar u = 
\int^u_{u^*}  \exp\left[ - \kappa_* (\bar u - u_*) \right] \;  \exp\left[  - \int^{\bar u}_{u_*} \delta \kappa(\tilde u) \; d\tilde u \right] d \bar u. 
\label{E:eee}
\end{equation}
Here we are interested in situations in which the second exponential on the RHS is in some suitable sense (to be more carefully defined below) ``close to unity''.

\subsection{Introducing the approximation}

We now begin the approximation procedure: Let us now suppose that
\begin{equation}
\label{E:condition}
\left|\int^{u}_{u_*} \delta \kappa(\tilde u) \; d\tilde u \right| \leq  \epsilon^2 \ll 1,
\end{equation}
where we shall soon check the conditions under which this happens. Under this hypothesis we can re-write the exact result (\ref{E:eee}) as 
\begin{equation}
\int^u_{u^*}   \exp\left[  - \int^{\bar u}_{u_*} \kappa(\tilde u) \; d\tilde u \right] d \bar u = 
\int^u_{u^*}  \exp\left[ - \kappa_* (\bar u - u_*) \right] \; \left\{ 1 + O(\epsilon^2) \right\} d \bar u, 
\end{equation}
which we can integrate to yield
\begin{equation}
\int^u_{u^*}   \exp\left[  - \int^{\bar u}_{u_*} \kappa(\tilde u) \; d\tilde u \right] d \bar u =
\left\{ {1- \exp\left[ - \kappa_* ( u - u_*) \right]  \over \kappa_*} \right\}  + O(\epsilon^2).
\end{equation}
The analysis here is somewhat delicate, because one is integrating a small quantity over what could be a very long time.
Note that the way we have set things up, this approximation will always be valid over \emph{some} interval --- the only real question is how long this validity interval will be. (See section \ref{S:range} below.)

The net result of the discussion up to this point is that, if we accept the condition (\ref{E:condition}), 
then for $u$ sufficiently close to $u_*$ we can effectively replace $\kappa(u)$ by $\kappa_*$,  and so write
\begin{eqnarray}
U 
&\approx&
U_* 
+ C_* \;  \int^{u}_{u^*} 
 \exp\left[  -\kappa_* \; (\bar u-u_*)  \right]
d\bar u
\nonumber
\\
&=&
U_* 
- {C_*\over\kappa_*}  \;  \left\{ 
 \exp\left[  -\kappa_* \; (u-u_*)  \right]
- 1 \right\}
\nonumber
\\
&=&
\left\{ U_* + {C_* \over \kappa_*} \right\}
- \left\{   {C_*\over\kappa_*}  \; e^{\kappa_*\; u_*}  \right\}
\;  \exp\left(  - \kappa_* u \right)
\nonumber
\\
&=&
U_H^*
- A_* \;  \exp\left(  - \kappa_* u \right),
\label{E:exp}
\end{eqnarray}
where we have defined
\begin{equation}
U_H^* = u_* + {C_*\over \kappa_*}
\qquad \hbox{and} \qquad
A_* =  {C_*\over\kappa_*} \; e^{\kappa_*\; u_*}.
\end{equation}
In spite of the similarity with Hawking's approximation, it is vitally important to note that $U_H^*$ is \emph{not} the location of the horizon (extrapolated back to $\scri^-$) --- it is instead the \emph{best estimate} (based on what you can see locally at $u_*$) of where a horizon \emph{might be likely to form} if the relation between $U$ and $u$ keeps $e$-folding in the way it is at $u_*$. There is no actual implication that a strict event horizon (or indeed any sort of horizon) ever forms, only that it ``looks like'' a horizon might form in the ``not too distant future''.
Once we have this approximate relation, 
\begin{equation}
U = p(u)\approx U_H^*
- A_* \;  \exp\left(  - \kappa_* u \right),
\label{E:approx-exp}
\end{equation}
which we shall refer to as the ``exponential approximation'', then
the rest of the calculation simply  drops out (in the quite usual manner).
The only tricky point lies in estimating the range of validity of this ``exponential approximation''.

\subsection{Range of validity of the approximation}
\label{S:range}

The exponential approximation condition (\ref{E:condition}), can always be satisfied for small-enough integration intervals $(u-u_*)$. But how small is small-enough? 
As a mild technical assumption, let us consider only functions $\kappa$ such that we can define a constant
\begin{equation}
D:= \sup_{n>0}\left[{1 \over (n+1)!}\;{|\kappa^{(n)}_*| \over \kappa^{n+1}_*}\right]^{1/(n+1)}, \qquad D<+\infty.
\label{sup}
\end{equation}
Physically this amounts to the assumption that the only two relevant (reciprocal) timescales in the problem are $\kappa_*$ and $D\kappa_*$. (Any other scale is assumed to be smaller, which is simply another way of saying that $\kappa(u)$ is slowly varying over the region of interest.) This condition is rather mild, covering even functions with poles at specific values of $u$.  So $\kappa(u)$ is even allowed to exhibit ``sudden singularities'', with this terminology being borrowed from cosmology~\cite{barrow, barrow2, cattoen}.

Under this hypothesis we have
\begin{eqnarray}
\left|\int^{u}_{u_*} \delta \kappa(\tilde u) \; d\tilde u \right| &=&
\left| \sum_{n=1}^{+\infty} {1 \over (n+1)!}\kappa^{(n)}_* \; (u-u_*)^{n+1}\right|
\nonumber
\\
&\leq&
\sum_{n=1}^{+\infty} {1 \over (n+1)!}|\kappa^{(n)}_*| \; |u-u_*|^{n+1}
\nonumber
\\
&\leq& 
\sum_{n=1}^{+\infty} D^{n+1}\kappa^{n+1}_*|u-u_*|^{n+1}.
\label{exp-exp-1}
\end{eqnarray}
Let us temporarily set $x = D\kappa_*|u-u_*|$, then
\begin{equation}
\left|\int^{u}_{u_*} \delta \kappa(\tilde u) \; d\tilde u \right| \leq \sum_{n=2}^{+\infty} x^n= {1 \over 1-x}-1-x = {x^2\over 1-x}.
\end{equation}
Now, as long as 
\begin{equation}
{x^2\over 1-x} \leq \epsilon^2,
\end{equation}
we are sure that condition~(\ref{E:condition}) is satisfied. Taking into account that $\epsilon \ll 1$,  the previous condition is certainly guaranteed to hold as long as $x^2 \leq \epsilon^2/2$. (This is not the optimal condition, but it is simple and quite good enough for our purposes.)
Then 
\begin{equation}
2D^2\kappa^2_* \; (u - u_*)^2 \leq \epsilon^2 \ll 1.
\end{equation}
Thus the range of validity of the exponential approximation condition (\ref{E:condition}) is certainly at least as large as
\begin{equation}
|u-u_*| \leq {\epsilon  \over \sqrt{2}D\kappa_*} \ll {1  \over \sqrt{2}D\kappa_*}.
\end{equation}
In the most simple situations the first term in the definition~(\ref{sup}) dominates, thereby yielding
\begin{equation}
2D^2={|\dot \kappa_*| \over \kappa^2_*}. 
\end{equation}
For instance, for an evaporating Schwarzschild black hole one can estimate
\begin{equation}
|\kappa^{(n)}| \sim  
1\times 4\times 7\times \cdots\times (3n-2)\; 
{M_\mathrm{Planck}^{4n+2}\over M^{3n+1}},
\end{equation}
and so
\begin{eqnarray}
\left[{1 \over (n+1)!}\;{|\kappa^{(n)}_*| \over \kappa^{n+1}_*}\right]^{1/(n+1)} 
\sim  
\left[ {1\over n+1} \; {M_\mathrm{Planck}^{2n}\over M^{2n}}\right]^{1/(n+1)} 
\sim
{M_\mathrm{Planck}^2\over M^2} \; \left[ {M\over M_\mathrm{Planck}}\right]^{2/(n+1)}_. 
\qquad
\end{eqnarray}
It is the trailing factor of $(M/M_\mathrm{Planck})^{2/(n+1)}$ that guarantees that for $M\gg M_\mathrm{Planck}$ the lowest derivative dominates in the definition of $D$. In fact careful estimates putting in all relevant numerical pre-factors show that for a Schwarzschild black hole the first term (the $\dot \kappa$ term) dominates down to some 5 Planck times before the final evaporation event.

In typical situations like the above, where the $\dot\kappa$ term dominates in the definition of $D$, the exponential approximation holds for an interval at least as large as
\begin{equation}
|u-u_*| \leq \epsilon  |\dot \kappa_*|^{-1/2} \ll |\dot \kappa_*|^{-1/2}.
\end{equation}
Hereafter we will assume that this is the case.
(Other more complicated situations can be dealt with in a tedious but straightforward manner.)

\smallskip
\noindent
What is the physics behind this condition?
\begin{itemize}
\item If one looks at the spacetime for a limited interval $\Delta u = |u - u_*|$ then the lowest possible frequency one can hope to resolve is $\Omega_\mathrm{min} \sim {1/\Delta u}$.
\item The condition $ |\dot \kappa_*| \; (u - u_*)^2 \leq \epsilon^2 \ll 1$ then collapses to the statement that over the time interval of interest $\Delta \kappa\lesssim \epsilon^2 \Omega_\mathrm{min} \ll \Omega_\mathrm{min}$. 
\item
That is, the ``exponential approximation'' is guaranteed to hold as long as the change in surface gravity is less than the minimum detectable frequency. 
\end{itemize}
Over what range of situations does this condition hold for macroscopic black holes?
For Schwarzschild black holes our previous estimate of $|\dot M|$ implies
\begin{equation}
\dot \kappa \sim {M_\mathrm{Planck}^6 \over M^4},
\end{equation}
so the condition for guaranteed validity of the exponential approximation reduces to
\begin{equation}
|u-u_*| \ll  t_\mathrm{Planck}  \times \left({M\over M_\mathrm{Planck}}\right)^2
\sim 10^{15} \times (\hbox{age of universe})  \times \left({M\over M_\sun}\right)^2.
\end{equation}
So for stellar mass black holes there is absolutely no difficulty in satisfying this inequality.

Note that we can rewrite the condition
\begin{equation}
|\dot \kappa_*| \; (u - u_*)^2 \leq \epsilon^2 \ll 1,
\end{equation}
as 
\begin{equation}
{|\dot \kappa_*|\over \kappa_*^2}  \; \left[ \kappa_* (u - u_*)\right]^2 \leq \epsilon^2 \ll 1.
\end{equation}
Since we do want to be able to have $\kappa_* \; |u - u_*| \gg 1$, (that is, as we have just seen, macroscopic mass black holes should be analyzable via quasi-static techniques for many light crossing times), the condition for validity of the exponential approximation in turn implies
\begin{equation}
 {\dot \kappa_*\over \kappa_*^2} \ll \epsilon^2; \qquad \hbox{whence} \qquad  {\dot \kappa_*\over \kappa_*^2} \ll\!\!\ll 1. 
\end{equation}
Note now $|\dot \kappa_*/\kappa_*^2|$ is ``very very much less than 1''.  In this sense validity of the exponential approximation implies the adiabaticity condition.

\noindent
It is now useful to define an interval ${\cal S}_+$ in terms of the affine parameter on $\scri^+$,
\begin{equation}
{\cal S}_+ = \left(u_* - {\epsilon|\dot \kappa_*|^{-1/2}},  \; u_* + {\epsilon|\dot \kappa_*|^{-1/2}}\right),
\end{equation}
and the corresponding interval ${\cal S}_-$ defined in terms of the affine parameter on $\scri^-$,
\begin{equation}
{\cal S}_- =  p({\cal S}_+) = \left(U_H^* - {C_*\over\kappa_*}\;\exp\left[ {\epsilon\,\kappa_*|\dot\kappa_*|^{-1/2}}\right],  \; 
U_H^* - {C_*\over\kappa_*}\;\exp\left[ -{\epsilon\,\kappa_*|\dot\kappa_*|^{-1/2}}\right]\; \right).
\end{equation}
The ``exponential approximation'' is then valid for $u\in {\cal S}_+$, which is equivalent to $U\in {\cal S}_-$.
Note that for $\sqrt{|\dot\kappa_*|}/\kappa_*$ sufficiently small we can get ``arbitrarily close'' to the ``\emph{might be horizon}'' at $U_H^*$ while still remaining within the realm of validity of the exponential approximation.

\subsection{Nonlocal normalization}

To determine the constant $C_*$, and so also ultimately determine the constant $A_*$  appearing in the exponential approximation,  requires an overall choice of normalization --- we need to relate the overall scale of the affine coordinate on $\scri^-$ to that for $\scri^+$. A  natural choice is to demand
\begin{equation}
p(u\to-\infty) \to u.
\end{equation}
This has the effect of making sure that in the infinite past (\ie, before any collapse or other dynamics was initiated) $\scri^-$ and $\scri^+$ are connected in the simple sensible way
\begin{equation}
{d U\over du} \to 1 \qquad \hbox{as} \qquad u \to -\infty.
\end{equation}
But from
\begin{equation}
{dU\over du} =  
C_* \;    
\exp\left[  -\int^{u}_{u_*} \kappa(\tilde u) \; d\tilde u \right],
\end{equation}
we then see
\begin{equation}
C_* =
\exp\left[  -\int_{-\infty}^{u_*} \kappa(\tilde u) \; d\tilde u \right],
\end{equation}
So we see that $C_*$ depends on the entire past history of $\scri^+$ (the history of the collapse in Hawking's language).  Furthermore we now estimate that a horizon might form at
\begin{equation}
U_H(u_*)  = U_H^* = U_* +
{1\over \kappa_*} \; 
\exp\left[  -\int_{-\infty}^{u_*} \kappa(\tilde u) \; d\tilde u \right],
\end{equation}
and that
\begin{equation}
A_* = 
{1 \over \kappa_*} \; 
\exp\left[  -\int_{-\infty}^{u_*} \kappa(\tilde u) \; d\tilde u +\kappa_* \; u_*\right].
\end{equation}
So finally we have explicit formulae for the parameters appearing in the exponential approximation. 

We now also have a clear physical interpretation for $C_*$: Since 
\begin{equation}
\left.{d U\over d u}\right|_{u_*} = C_*,
\end{equation}
we see that a small time interval $\Delta U$ around $U_*$ on $\scri^-$ is related to a small time interval $\Delta u$ around $u_*$ on $\scri^+$ by
\begin{equation}
\label{E:redshift}
\Delta U = C_* \; \Delta u.
\end{equation}
This implies that $C_*$ is the Doppler shift factor (redshift factor) relating the frequency of a photon emitted at $U_*$ on $\scri^-$ to its frequency when it reaches $u_*$ on $\scri^+$.

\subsection{Logarithmic  approximation}

One can also invert the exponential relation (\ref{E:approx-exp}) to give the perhaps more common ``logarithmic approximation'' ~\cite{hawking1,hawking2}
\begin{equation}
u \approx  - {1\over\kappa}\;\ln\left\{\; {U_H^*-U\over A_*}\;\right\}; 
\qquad\qquad U\in {\cal S}_-.
\end{equation}
For completeness, we note that in terms of the formalism we have developed above, the definition
\begin{equation}
\kappa(u) = \left. -{d^2 U/du^2\over {dU/du} }\right.,
\end{equation}
can, after some manipulations, be seen to be equivalent to
\begin{equation}
\kappa(U) = \left. {d^2 u/dU^2\over (du/dU)^2 }\right.,
\end{equation}
which can be integrated as
\begin{equation}
u(U) = u_* + \int_{U_*}^U {d \bar U\over C_* - \int_{U_*}^{\bar U} \kappa(\tilde U) d \tilde U},
\end{equation}
with the same $C_*$ as occurred previously. This can be used as a basis for an alternative way of deriving the ``logarithmic approximation'' above. 

\section{Peeling properties versus inaffinities}

To obtain a clear physical interpretation for the function $\kappa(u)$, we will now connect the behaviour of the affine null parameters $u$ and $U$ back to the geometry as encoded in the spacetime metric. The most efficient way of doing so is by adopting double null coordinates $\{W,u\}$ on the transverse slice (the $\{t,r\}$ sector). We use $W$ as the null coordinate rather than $U$, to emphasise that $W$ and $u$ are to be viewed as independent coordinates on the spacetime, whereas $U=p(u)$ and $u$ are linked by being different labels for the same null curve (reflected through the origin). We wish specifically to clarify the relationship between $\kappa(u)$ and the ``peeling properties'' of outgoing null geodesics, and contrast this with the inaffinity properties of null geodesics on the horizon. See the appendix for additional background. (See also~\cite{njp} and~\cite{Macher} for comments on the importance of these ``peeling'' properties in ``analogue spacetimes, and~\cite{Brout1, Brout2} for comments in a general relativistic context.)

\subsection{Metric asymptotics}

For any spherically symmetric spacetime, we can without loss of generality adopt such double-null coordinates and set
\begin{equation}
d s^2 = - F(W,u) \;d W \, d u + r(W,u)^2 \;d\Omega^2_{d-1}.
\end{equation}
See, for instance, Bergmann and Roman~\cite{bergmann-roman}. Here the two functions $F(W,u)$ and $r(W,u)$ completely specify the spacetime geometry.
Since we have chosen $W$ to be affine on $\scri^-$ and $u$ to be affine on $\scri^+$ we can impose
\begin{equation}
 \lim_{W\to+\infty} F(W,u) = 1; \qquad\qquad \lim_{u\to-\infty} F(W,u) = 1.
 \end{equation}
 This makes the metric particularly simple on $\scri$.
 
\subsection{Inaffinity estimation in the bulk}

Recall that the generators of $\scri^-$ are null geodesics, affinely parameterized by $W$. Using $(W, u, \theta_1, \dots, \theta_{d-1})$ coordinates, throughout the spacetime one can define the null vector field
\begin{equation}
k^a = (1,0,0,\cdots,0),
\end{equation}
pointing in the direction of increasing $W$. For that null geodesic that skims along $\scri^-$ this is an affine parameter, for other null curves (labelled by a constant value of $u$) which traverse the interior of the spacetime this is no longer an affine parameter. The 4-acceleration
\begin{equation}
k^a \nabla_a k^b 
= \Gamma^b{}_{WW} 
= {F_{,W}\over F} (1,0,\cdots,0),
\end{equation}
now enables us to identify a new and logically distinct quantity
\begin{equation}
\kappa_\mathrm{bulk}(W,u) = {F_{,W}\over F}.
\end{equation}
This is now a ``bulk'' quantity defined everywhere in the spacetime, (not just on horizons or asymptotic infinity). This $\kappa_\mathrm{bulk}$ is a ``bulk inaffinity estimator'' which measures the extent to which the coordinate $W$ fails to be an affine parameter along null geodesics  of increasing $W$. It is this ``bulk inaffinity estimator''  that is closely related to the textbook notion of surface gravity, while it is the ``peeling'' notion of surface gravity, $\kappa(u)$, that we have seen is related to the Hawking flux. 

\subsection{The surface gravity of ${\cal H}^+$}

If (and this is one of Hawking's key assumptions in his derivation of the existence of a Planck flux reaching $\scri^+$) a true future-eternal event horizon forms, then there will be a region of $\scri^-$ (namely $W>U_H$) which has no natural ``lift'' to $\scri^+$. Instead, this region of $\scri^-$ ``lifts'' to ${\cal H}^+$, the future-eternal event horizon.  On ${\cal H}^+$ we have an (in principle evolving $W$-dependent) inaffinity notion of surface gravity: 
\begin{equation}
\kappa_\mathrm{inaffinity}(W)  = \lim_{u\to+\infty}  {\kappa_\mathrm{bulk}(W,u)}.
\end{equation}
Note particularly that by construction the textbook notion of surface gravity makes sense only on the future event horizon ${\cal H}^+$. In contrast, note that our definition $\kappa(u)$ naturally resides on $\scri^+$, not on ${\cal H}^+$.

It is only if we make the further assumption that, after the future-eternal  event horizon forms, the black hole settles down to an asymptotically static state, that we have a relation
\begin{equation}
\kappa_H = \lim_{W\to+\infty} \kappa_\mathrm{inaffinity}(W) =  \lim_{u\to+\infty} \kappa(u),
\end{equation}
where $\kappa_H$ is now the quantity appearing in Hawking's original derivation.
We emphasise that without such an asymptotic assumption the two notions (inaffinity \emph{vs} peeling) are distinct, and in general different. Even in this particular case they at best only coincide at $i^+$, future timelike infinity, and are unrelated at other locations. It is only for static (or stationary) spacetimes that the two notions exactly coincide for all times. (See appendix \ref{A:2}).

If in contrast one has a future non-eternal event horizon, (an event horizon that forms and subsequently evaporates  as described by the standard conformal diagram for an evaporating black hole, see for instance~\cite{lindesay+sheldon}), then as seen from $\scri^+$ the horizon both forms and evaporates ``simultaneously'' at some $u_H<+\infty$, and one has
\begin{equation}
U_H = p(u_H^-),  \qquad  U_E = p(u_H^+),
\end{equation}
where $U_E$ now labels the first incoming null ray that reaches the centre \emph{after} the evaporation process is complete.  Note that at the end-point $u_H$ of the evaporation $p(u)$ has a discontinuity, and $\kappa(u)$ will diverge. 
This strongly suggests that in this situation some form of ``thunderbolt'' will be emitted~\cite{thunderbolt, thunderbolt2}.

\subsection{Summary}

The physics here is intriguing~\cite{our-prl-article} --- to get a Planck flux not only do we not ever need the future horizon ${\cal H}^+$ to form, but the Hawking temperature is not logically or physically connected to the surface gravity of the horizon ${\cal H}^+$. Instead the Hawking temperature is primarily related to $\kappa(u)$, that is, the ``peeling off'' properties of the null geodesics that actually do reach $\scri^+$,  and hence to the peeling notion of surface gravity.
It is only when a collapse settles down to a static black hole spacetime (the end point of classical collapse in general relativity) that these quantities happen to be equal.

\section{Bogoliubov coefficients: Basic framework}

The Bogoliubov coefficient calculation --- which proceeds from the exponential approximation to the Hawking flux --- is in principle completely standard~\cite{hawking1, hawking2, birrell-davies}. Modulo normalizations and technical issues, such a calculation will automatically give us a steady Planck flux of particles at $\scri^+$ as long as the exponential approximation holds for a sufficiently large interval. The purpose of presenting the calculation in some detail is to provide a very general and clean derivation, exhibit some novel results using the stationary phase technique for evaluating the relevant integrals, and to pull together a few  tricks otherwise scattered throughout the literature.

To place this discussion in a textbook perspective: Note that the ultimate goal is to reproduce some close analogue of the discussion in  Birrell \& Davies~\cite{birrell-davies}; either equations (4.59), (4.60), and (4.61) on page 108 of the ``moving mirror" discussion of pages 102--109, or the closely related equation (8.32) in the collapsing black hole discussion on pages 250--260, but now with slowly varying time-dependent Bogoliubov coefficients.

\subsection{Bogoliubov coefficients in terms of the Klein--Gordon inner product}

It is very easy to write down the asymptotic modes for a massless scalar field.
On $\scri^-$ the properly normalized modes are (see for instance Birrell \& Davies equation (8.29) and (8.30) on page 258; note we are now in 3+1 dimensions)
\begin{equation}
{Y_{\ell m}\over\sqrt{4\pi\omega}\;r} \;\exp(-i\omega U).
\end{equation}
These modes extend to the full spacetime as
$\phi^U(\omega; t,\vec x)$. 
Similarly, on $\scri^+$ the properly normalized asymptotic modes are
\begin{equation}
{Y_{\ell m}\over\sqrt{4\pi\omega}\;r} \; \exp(-i\omega u),
\end{equation}
which extend to the full spacetime as
$\phi^u(\omega;t,\vec x)$. 
The \emph{exact} Bogolubov coefficients are now easily written down (in terms of the Lorentz-invariant Klein--Gordon inner product, see Birrell \& Davies equation (2.9) on page 11, and equation (3.36) on page 46) as
\begin{eqnarray}
\alpha(\omega,\omega') &=&  ( \phi^u(\omega;t,\vec x)  ,  \phi^U(\omega';t,\vec x)   )
\\
&=& -i \int d^3 x 
\left\{ 
\phi^u(\omega;t,\vec x) \; \partial_t \phi^U(\omega';t,x)^* -
 \phi^U(\omega';t,\vec x)^* \; \partial_t \phi^u(\omega;t,x)
\right\},
\nonumber
\end{eqnarray}
\begin{eqnarray}
\beta(\omega,\omega') &=&  -( \phi^u(\omega;t,\vec x)  ,  \phi^U(\omega';t,\vec x)^*   )
\\
&=& -i \int d^3 x 
\left\{ 
\phi^U(\omega';t,\vec x) \; \partial_t \phi^u(\omega;t,x) -
 \phi^u(\omega;t,\vec x) \; \partial_t \phi^U(\omega';t,x) 
\right\},
\nonumber
\end{eqnarray}
where the integral runs over any arbitrary spacelike hypersurface that terminates at spacelike infinity (that is, at $i^0$).
Note that we associate $\omega'$ with the frequency of the mode that is simple on $\scri^-$, and $\omega$ with the mode that is simple on $\scri^+$.

Assuming for the moment that no event horizon forms, we now choose the spacelike hypersurface to skim arbitrarily close to $\scri^+$. (This is a slightly unusual proposal, because calculations in the literature, where one commonly assumes that an event horizon forms, are normally performed on $\scri^-$~\cite{hawking1, hawking2, birrell-davies}.
However, if no horizon forms, the ultimate results are equivalent, because the Bogoliubov coefficient, being a Lorentz invariant  ``inner product", will not depend on which hypersurface one picks. Thus working on $\scri^+$ is a useful consistency check on the formalism. In particular, this way of organizing things emphasises the fact that actual horizon formation is not central to the Hawking process.)
Focussing on the $\beta$ coefficient we can evaluate:
\begin{eqnarray}
\beta(\omega,\omega') &=&  {-i \delta_{\ell \ell'} \; \delta_{mm'} \over 2\pi\; \sqrt{4\omega \;\omega' }} \;
\\
&& \times
 \int_{-\infty}^{+\infty}  du
\left\{ \vphantom{\bigg|}
\exp[-i\omega u] \; \partial_u \exp[-i\omega' p(u)] -
\exp[-i\omega' p(u)] \; \partial_u\exp[-i\omega u] 
\right\}.
\nonumber
\end{eqnarray}
Integrate the first term by parts. There will be a surface term proportional to
\begin{equation}
\left. \exp[-i\omega u] \exp[-i\omega'  \, p(u)]  \right|^{+\infty}_{-\infty},
\end{equation}
which can safely be discarded. The remaining term, suppressing the angular Kronecker deltas,  gives
\begin{equation}
\beta(\omega,\omega') ={1\over2\pi}\; \sqrt{\omega \over\omega'} \;
 \int_{-\infty}^{+\infty} du
\left\{ 
\exp[-i\omega u] \exp[-i\omega' p(u)]  \; 
\right\}.
\end{equation}
Compare with  Birrell \& Davies equation (4.59) on page 108. 
\emph{We emphasize that this formula, though it looks extremely simple, is in fact one of the key results needed in the derivation of the Hawking flux.}
We can also perform a change of variables in the above, and after an integration by parts obtain the completely equivalent formula
\begin{equation}
\label{E:beta-scri}
\beta(\omega,\omega') ={1\over2\pi}\; \sqrt{\omega' \over\omega} \;
 \int_{-\infty}^{+\infty} dU
\left\{ 
\exp[-i\omega \, p^{-1}(U)] \exp[-i\omega' U]  \; 
\right\}.
\end{equation}
Of course this is also exactly what you would get by evaluating the Bogoliubov coefficient  by using a spacelike hypersurface that skims just above $\scri^-$.

If in contrast one assumes that an event horizon does form, then it is certainly more efficient to work on $\scri^-$. (Since $\scri^-$ is always a complete Cauchy hypersurface, whereas if a horizon forms one needs to work with $\scri^+\cup{\cal H}^+$ to get a complete Cauchy hypersurface.) In any case there is universal agreement on the applicability of equation (\ref{E:beta-scri}). See for instance~\cite{hawking1, hawking2, birrell-davies}.

\subsection{Approximation scheme}

The novelty in the current analysis is this: Expand the \emph{integrand} around $U=U_*$, corresponding to
$u=u_*$, where we have already seen 
that we can justify the ``exponential approximation'' 
\begin{equation}
U \approx U_H^* - A_* \;  \exp(  - \kappa_*u).
\end{equation}
In view of this, we have the approximate result
\begin{equation}
\exp[-i\omega' U] = \exp[-i\omega' p(u)] \approx \exp[-i\omega' U_H^*]  \; \exp\left\{ +i\omega' A_*  \exp[-\kappa_* u]\right\},
\end{equation}
which we know to be valid in the range
\begin{equation}
|u-u_*|  \leq {\epsilon |\dot\kappa_*|^{-1/2}}.
\end{equation}
Then expanding the integrand (not the integral)
\begin{eqnarray}
&&\beta(\omega,\omega') \approx {1\over2\pi}\; \sqrt{\omega \over\omega'} \;
\exp[-i\omega' U_H^*] \;
\nonumber
\\
&&\qquad
\times 
\int_{u_*-{\epsilon|\dot\kappa_*|^{-1/2}}}^{u_*+{\epsilon|\dot\kappa_*|^{-1/2}}}  du \;
\exp[-i\omega u] \exp[+i\omega' A_* \exp\{-\kappa_* u \}].
\end{eqnarray}
The most significant pieces we are currently neglecting in this approximation are the contribution from the ``tail'' integration regions 
\begin{equation}
\label{E:tail}
\left(-\infty, \; u_*-{\epsilon|\dot\kappa_*|^{-1/2}}\right) \qquad  \hbox{and} \qquad  \left(u_*+{\epsilon|\dot\kappa_*|^{-1/2}}, \; +\infty\right).
\end{equation}
That is, we are neglecting the contribution from the region $I\!\!R - {\cal S}_+$, since in the region
\begin{equation}
|u-u_*|  \leq {\epsilon|\dot\kappa_*|^{-1/2}} \qquad \Leftrightarrow \qquad u \in  {\cal S}_+
\end{equation}
we \emph{are} accurately estimating the value of the integrand. But in the tail regions the phase is rapidly oscillating, so the tail contribution to the Bogoliubov coefficient is small.

Hawking's key approximation~\cite{hawking1, hawking2}  (suitably rephrased, reinterpreted, and now extended to slowly evolving systems) is this: In view of the condition underlying the exponential approximation, one is justified in extending this approximation
for the integrand to all $u$, but \emph{provided that one then interprets the
  resulting Bogoliubov coefficient as the ``effective'' $\beta(u_*)$
  for the production of particles (wave packets) that arrive at $\scri^+$ at $u\approx
  u_*$}.  Within the context of this approximation we now have
\begin{equation}
\beta(u_*;\omega,\omega') \approx 
{1\over2\pi}\;
\sqrt{\omega \over\omega'} \;
\exp[-i\omega' U_H^*] \;
 \int_{-\infty}^{+\infty} du\;
\exp[-i\omega u +i\omega' A_*  \exp\{-\kappa_* u \}].
\end{equation}
One is effectively assuming that the true integrand and approximate integrand are both rapidly oscillating in the ``tail'' integration regions in (\ref{E:tail}), 
and so both average to zero on these intervals.

We emphasize that this approximation (though without the extension to time-dependent Bogoliubov coefficients)  is actually what Hawking implicitly does in his original 1974
calculation, with slight modifications since he is working on $\scri^-$, and for the particular case $U_H^* = U_H$, corresponding to formation of a strict event horizon where 
$u_*\to+\infty$. In Hawking's \emph{Nature} article~\cite{hawking1}, the approximation is hidden in the third column where he first writes down the \emph{approximate} asymptotic form of the mode and then performs an \emph{exact} Fourier transform for this asymptotic \emph{approximation}.  Hawking makes essentially the same approximation in his \emph{Communications in Mathematical Physics} article~\cite{hawking2}, where (2.18) is his asymptotic \emph{estimate} for the mode on $\scri^-$, and by (2.19) he has performed an \emph{exact}\/ Fourier transform of  this \emph{approximate} asymptotic mode.

One can also justify the approximation using a wave-packet argument. 
We know the exponential approximation with fixed $u_*$ is valid for $u\in {\cal S}_+$. Now consider some wave-packet that arrives at $\scri^+$ with compact support in ${\cal S}_+$.  Physically, such a wave packet cannot tell the difference between the (unkown) exact relation $u = p^{-1}(U)$, the ``exponential approximation" defined above,  and Hawking's (at first glance seemingly absurd) extrapolation of the exponential approximation to ``all time".
That is, for any wave-packet with compact support on ${\cal S}_+$, where by definition the quantity $\kappa(u)\approx\kappa(u_*)$ is approximately constant, we can use the standard Hawking calculation to derive the Bogoliubov coefficients relevant to wave-packets confined to this time slice --- that is, $\beta(u_*; \omega,\omega')$. 
We see that  \emph{up to an irrelevant overall phase}:
\begin{equation}
\beta(u_*;\omega,\omega') \approx 
{1\over2\pi}\;
{\sqrt{\omega \over \omega'}} \;
  \int_{-\infty}^{+\infty} du\;
\exp[-i\omega u +i\omega' A_*  \; \exp\{-\kappa_*\; u\}].
\end{equation}
At this stage you can see that the standard Hawking result is really
unavoidable. 
To complete the calculation we can now \emph{either} approximate this integral using stationary phase techniques,
\emph{or} go for an ``exact'' evaluation using Gamma functions.

\section{Stationary phase evaluation of the Bogoliubov coefficients}

Stationary phase approximations have been somewhat neglected in this part of the technical literature, but appear to be at least as reliable as the usual ``exact'' results obtained via Gamma function identities. One actually has to do two slightly different but very closely related calculations --- a Boltzmann calculation to get the overall normalization and  a calculation of the ratio $|\beta/\alpha|$ to obtain a Planck spectrum.

\subsection{Boltzmann spectrum using stationary phase}

The stationary phase approximation is
\begin{equation}
\int_{-\infty}^{+\infty} \exp[-i \phi(u)] d u \approx 
\sqrt{2\pi i \over\phi''(u_s)} \; \exp[-i \phi(u_s)],
\end{equation}
where $u_s$ is defined by $\phi'(u_s)=0$. This represents the leading term in an asymptotic approximation based on a Gaussian expansion of the integrand around its stationary point.
In the current situation the relevant phase is
\begin{equation}
\phi(u) = \omega  u - \omega' A_*   \exp(-\kappa_* u),
\end{equation}
and the stationary phase condition is given by
\begin{equation}
\phi'(u_s) = \omega + \omega' A_* \kappa_*  \exp(-\kappa_* u_s) = 0.
\end{equation}
That is 
\begin{equation}
 \exp(-\kappa_* u_s) = - {\omega\over\omega' A_* \kappa_*},
\end{equation}
whence
\begin{equation}
u_s = -{1\over\kappa_*} 
\ln\left( {\omega\over\omega' A_* \kappa_*} \right) - {i \pi \over\kappa_*}.
\end{equation}
Note that $u_s$, the location of the ``stationary'' phase, is
\emph{complex}.  
Now 
\begin{equation}
\phi''(u_s) = -\omega'  A_* \kappa_*^2 \exp(-\kappa_* u_s)  =  \kappa_* \; \omega.
\end{equation}
The stationary phase approximation is now
\begin{equation}
\beta(u_*;\omega,\omega') \approx 
{1\over2\pi} \;
\sqrt{\omega \over\omega'} \;  \sqrt{2\pi i\over \kappa_* \omega} \;
\exp[-i\omega u_s +i\omega' A_*  \exp(-\kappa_* u_s)],
\end{equation}
whence
\begin{equation}
\beta(u_*;\omega,\omega') \approx 
 {1\over \sqrt{2\pi\kappa_* \omega'} }\;
 \exp\left[-i\omega u_s -i{\omega\over \kappa_*}+i{\pi\over4}\right].
\end{equation}
The only really interesting bit comes from the \emph{imaginary} part of $u_s$
which yields, \emph{up to a physically irrelevant overall phase}:
\begin{equation}
\beta(u_*;\omega,\omega') 
\approx {1\over \sqrt{2\pi\kappa_* \omega'} }\;
\exp\left[-{\omega\pi\over\kappa_*}\right].
\end{equation}
This gives
\begin{equation}
|\beta(u_*;\omega,\omega')|^2 
\approx {1\over 2\pi\kappa_* \omega' }\;
 \exp\left[-{2\pi\omega\over\kappa_*}\right],
\end{equation}
which is a Boltzmann factor.
Compare the normalization with  Birrell \& Davies equation (4.61) on page 108.
Note that this elementary application of the stationary phase approximation does not (yet) yield a
  Planck spectrum, though it does yield a Boltzmann spectrum, and one quickly gets to the Hawking temperature this way:
\begin{equation}
k_B \;T_H(u_*) = {\kappa(u_*)\over2\pi}= {\kappa_*\over2\pi}.
\end{equation}
Note that the size of the sub-leading terms in the asymptotic expansion underlying the stationary phase approximation will be controlled by the width of the Gaussian that is its leading term, that is by $\phi''(u_s)=\omega \kappa_*$.  This implies that the stationary phase approximation (as we have used it so far) will be valid up to fractional corrections of order $O(\kappa_*/\omega)$.  This is why the current version of the stationary phase approximation is intrinsically a high-frequency approximation, incapable of seeing the sub-leading exponentials critical to probing the Planck nature of the spectrum.

\subsection{Planck spectrum using stationary phase}

To get a Planck (rather than Boltzmann) spectrum
use the \emph{exact} normalization condition
\begin{equation}
\label{E:norm}
\int d\omega' \left[\alpha(\omega_1,\omega')\alpha^*(\omega_2,\omega')-\beta(\omega_1,\omega')\beta^*(\omega_2,\omega')\right] =\delta(\omega_1-\omega_2),
\end{equation}
and consider the ratio
\begin{equation}
{\beta(u_*;\omega,\omega')\over\alpha^*(u_*;\omega,\omega')} 
\approx 
{
\int_{-\infty}^{+\infty} du
\exp[-i\omega u +i\omega' A_*  \; \exp(-\kappa_* u)]
\over
\int_{-\infty}^{+\infty} du
\exp[-i\omega u -i\omega' A_*  \; \exp(-\kappa_* u)]}.
\end{equation}
Now use stationary phase on numerator and denominator separately. We shall see that this ratio is much better behaved than each individual integral. The relevant phases and derivatives are:
\begin{equation}
\phi = \omega u \mp \omega' A_*  \; \exp(-\kappa_* u),
\end{equation}
\begin{equation}
\phi' = \omega \pm \omega' A_* \kappa_* \exp(-\kappa_* u),
\end{equation}
\begin{equation}
\phi'' = \mp \omega' A_* \kappa_*^2 \exp(-\kappa_* u).
\end{equation}
Furthermore for $n \geq 2$
\begin{equation}
\phi^{(n)} = \mp (-1)^n  \omega' A_* \kappa_*^n \exp(-\kappa_* u).
\end{equation}
The two stationary points differ \emph{only} in their imaginary part 
\begin{equation}
u_s = -{1\over\kappa_*} 
\ln\left( {\omega\over\omega' A_* \kappa_*} \right) - {\ln(\mp1)\over\kappa_*},
\end{equation}
which we can rewrite as
\begin{equation}
u_s = u_* -{1\over\kappa_*} 
\ln\left( {\omega\over\omega' C_* } \right) + \left\{ {i\pi\over\kappa_*}, 0 \right\}.
\end{equation}
For both integrals
\begin{equation}
\phi''(u_s) = \kappa_* \omega,
\end{equation}
and for $n\geq 2$
\begin{equation}
\phi^{(n)}(u_s) =  (-1)^n  \kappa_*^{n-1} \omega,
\end{equation}
while for the phases
\begin{equation}
\phi(u_s) = \omega u_s + \omega/\kappa_*.
\end{equation}
Since the two stationary points differ only in their imaginary part, we see that 
\begin{equation}
{\beta(u_*;\omega,\omega')\over\alpha^*(u_*;\omega,\omega')} 
\approx \exp(-\pi \omega/\kappa_*),
\end{equation}
and we are essentially done. Note that the sub-leading terms in the asymptotic expansion for $\beta$ are the same as the sub-leading terms in the asymptotic expansion for $\alpha$, so they quietly cancel, and the stationary phase approximation for the ratio $|\beta/\alpha|$ is much better behaved than the stationary phase approximation for each term individually. 
In fact, using the normalization found in the previous (Boltzmann) version of the stationary phase calculation and showing explicitly their phase factor we can write
\begin{eqnarray}
\beta(u_*;\omega,\omega') \approx  {1\over 2\pi \kappa_* \omega'} \exp\left[-i{\omega \over \kappa_*}
\ln\left({\omega' A_* \kappa_* \over \omega}\right)\right] \beta_\omega,
\\
\alpha^*(u_*;\omega,\omega') \approx  {1\over 2\pi \kappa_* \omega'} \exp\left[-i{\omega \over \kappa_*}
\ln\left({\omega' A_* \kappa_* \over \omega}\right)\right] \alpha_\omega,
\end{eqnarray} 
where $\alpha_\omega, \beta_\omega$ are real functions of $\omega$ such that $\beta_\omega/\alpha_\omega = \exp(-\pi \omega/\kappa_*)$. In this way, using~(\ref{E:norm}) it is easy to check that we now get
\begin{equation}
|\beta(u_*;\omega,\omega')|^2 \approx {1\over 2\pi \kappa_*\, \omega'} \;\;
{1\over \exp(2\pi \omega/\kappa_*) - 1},
\end{equation}
which is indeed a Planck spectrum.
Compare, for instance, with Birrell \& Davies equation (4.61), page 108.

\subsection{Mathematical range of validity for stationary phase}
A tricky point of the stationary phase approximation is this:  Is the location $u_s$
of the stationary 
point in the phase within the range of validity of the exponential approximation?
To check this consider 
\begin{equation}
\left| \mathrm{Re}(u_s-u_*)\right| = \left|-{1\over\kappa_*} 
\ln\left( {\omega\over\omega' A_* \kappa_*} \right)  - u_*\right|
= \left|{1\over\kappa_*} 
\ln\left( {\omega\over\omega' C_* } \right) \right|.
\end{equation}
Thus mathematically the stationary phase approximation will be valid as long as
\begin{equation}
 \left|\ln\left( {\omega\over\omega' C_*} \right) \right|  \leq {\epsilon\, \kappa_*|\dot\kappa_*|^{-1/2}},
 \label{E:range1}
\end{equation}
that is
\begin{equation}
\exp\left[  - {\epsilon\, \kappa_*|\dot\kappa_*|^{-1/2}}\right]
\leq  \quad {\omega\over C_* \omega'}  \quad
\leq  \exp\left[  +{\epsilon\, \kappa_*|\dot\kappa_*|^{-1/2}}\right].
\label{E:range2}
\end{equation}
Remember that $\omega'$ is the unobserved frequency on $\scri^-$ while $\omega$ is the physically relevant frequency on $\scri^+$.
Now consider the quantity $C_* \; \omega'$.
What is its physical interpretation? Consider the null curve labeled by $u_*$, and a small interval $\Delta u$ centered on $u_*$. Then on $\scri^-$ this interval corresponds, (recall (\ref{E:redshift})), to $\Delta U = C_*   \; \Delta u$, that is
 \begin{equation}
 {1\over \Delta u} = C_*  \; {1\over \Delta U}.
 \end{equation}
Therefore
 \begin{equation}
\omega'_+ = C_*\;\omega',
 \end{equation}
where $\omega'_+$ is the ``Doppler shifted'' value of $\omega'$ if we transport it to $\scri^+$ along the null curve labelled by $u_*$. (That is, $\omega'_+$ is the frequency on $\scri^+$ corresponding to $\omega'$ on $\scri^-$ at ``time'' $u_*$.) 
  Then the condition for validity of the stationary phase approximation is 
\begin{equation}
 \exp\left[  - {\epsilon\, \kappa_*|\dot\kappa_*|^{-1/2}}\right]
\leq  \quad {\omega\over\omega'_+}  \quad
\leq \exp\left[  + {\epsilon\, \kappa_*|\dot\kappa_*|^{-1/2}}\right].
\end{equation}
This can always be satisfied for an extremely broad range of frequencies.
Indeed, for a Schwarz\-schild black hole we have already seen $\kappa\sim1/M$ and $\dot\kappa \sim M_\mathrm{Planck}^2/M^4$, therefore
\begin{equation}
 \exp\left[  - {\epsilon\,  M\over M_\mathrm{Planck}}\right]
\leq  \quad {\omega\over\omega'_+}  \quad
\leq \exp\left[  + {\epsilon\, M\over M_\mathrm{Planck}}\right].
\end{equation}
Thus mathematically there is no risk of the validity of the stationary phase approximation causing additional problems --- the validity of the exponential approximation, the adiabatic condition, and the validity of the stationary phase approximation are all secure for macroscopic black holes.  If we wish to more physically relate $\omega'$ to $\omega$, (currently they are completely independent quantities), that can be done once we perform a wave-packet analysis. See section \ref{S:wave-packet}.

\section{Gamma function evaluation  of the Bogoliubov coefficients}

As an alternative way of evaluating the integral (and hence yet another consistency check) we now use Gamma function techniques. 
Note that the logic to be presented in this section is at first sight slightly odd, \emph{but completely standard}~\cite{hawking1, hawking2}. We are taking an \emph{approximate} value for the Bogoliubov coefficient and then \emph{exactly} evaluating the resulting integral.  The ultimate justification for this comes from looking at a wave-packet supported on ${\cal S}_+$ and noting that for this wave-packet one might as well extend the integral to ``all time".   Discarding an irrelevant overall phase, start with the integral
\begin{equation}
\beta(u_*;\omega,\omega') \approx 
{1\over2\pi}\;
\sqrt{\omega \over\omega'} \; 
\int_{-\infty}^{+\infty} du
\exp[-i\omega u +i\omega' A_*  \exp(-\kappa_* u)].
\end{equation}
Now perform a change of variables
\begin{equation}
z =  \exp(-\kappa_* u),
\end{equation}
so that
\begin{equation}
\beta(u_*;\omega,\omega') \approx {1\over2\pi}\;
\sqrt{\omega \over\omega'} \;  {1\over \kappa_*}\;
\int_0^{+\infty} dz \;
z^{i\omega/\kappa_*-1} \; \exp( i\omega' A_* z).
\end{equation}
Now take 
\begin{equation}
\bar z = -i\omega' A_* z,
\end{equation}
so after an appropriate change of contour
\begin{equation}
\beta(u_*;\omega,\omega') \approx 
{1\over2\pi}\;
\sqrt{\omega \over\omega'} \; {1\over \kappa_*}\;
 ( -i\omega' A_*)^{-i\omega/\kappa_*} \;  \int_0^{+\infty} d\bar z \;
\bar z^{i\omega/\kappa_*-1} \; \exp( -\bar z),
\end{equation}
whence
\begin{equation}
\beta(u_*;\omega,\omega') \approx 
{1\over2\pi}\;
\sqrt{\omega \over\omega'} \; {1\over \kappa_*}\;
( -i\omega' A_*)^{-i\omega/\kappa_*} \Gamma(i\omega/\kappa_*).
\end{equation}
So up to an irrelevant phase
\begin{equation}
\beta(u_*;\omega,\omega') \approx 
{1\over2\pi}\;
\sqrt{\omega \over\omega'} \;  {1\over \kappa_*}\;
( -i)^{-i\omega/\kappa_*} \;\;\Gamma(i\omega/\kappa_*)
\end{equation}
\begin{equation}
\qquad\qquad\qquad
= {1\over2\pi}\; \sqrt{\omega \over\omega'} \;  {1\over \kappa_*}\;
\exp[-\pi \omega/(2\kappa_*)] \;\; \Gamma(i\omega/\kappa_*).
\end{equation}
Using the properties of Gamma functions
\begin{equation}
|\Gamma(ix)|^2 = {\pi\over x \sinh(\pi x)},
\end{equation}
we have
\begin{equation}
|\beta(u_*;\omega,\omega')|^2 \approx {1\over(2\pi)^2} \;
{\omega \over\omega'} \; {1\over \kappa_*^2}\;
\pi \; {\exp(-\pi\omega/\kappa_* )\over \omega/\kappa_* \; \sinh(\pi\omega/\kappa_*)},
\end{equation}
so that
\begin{equation}
|\beta(u_*;\omega,\omega')|^2 \approx {1\over 2\pi \kappa_* \omega'} \;\;
{1\over \exp(2\pi \omega/\kappa_*) - 1}.
\end{equation}
Compare with Birrell \& Davies equation (4.61), page 108.
This is a Planck spectrum with 
\begin{equation}
k_B \; T_H(u_*) = {\kappa(u_*)\over2\pi} =  {\kappa_*\over2\pi}.
\end{equation}
And we are done. Note however, that this calculation is actually no more ``exact''  than the stationary phase approximation; one has just hidden the approximations elsewhere --- in particular, an approximate form for the mode function near $u_*$ has been promoted to all of $\scri^+$.
That is: Our approximate mode function should be a good approximation to the real mode function for
\begin{equation}
|u-u_*|  \leq {\epsilon|\dot \kappa_*|^{-1/2}}.
\end{equation}
We then assume this region ${\cal S}_+$ dominates the integral, so that formally we can integrate the approximate mode function all the way from $-\infty$ to $+\infty$. 
Of course, the condition that this region ${\cal S}_+$ dominates the integral is equivalent to requiring the validity of the stationary phase approximation, leading again to the constraint:
\begin{equation}
\exp\left[  - {\epsilon\, \kappa_*|\dot\kappa_*|^{-1/2}}\right]
\leq  \quad 
 {\omega\over\omega'_+} \quad
\leq  \exp\left[  + {\epsilon\, \kappa_*|\dot\kappa_*|^{-1/2}}\right].
\end{equation}

\section{Physical particle detection: Wave-packets}
\label{S:wave-packet}

When we say that we are calculating the number of particles of frequency $\omega$ arriving to the asymptotic region at time $u_*$, what we really mean by particle is a normalized wave packet with a frequency content concentrated around the frequency $\omega$ while simultaneously being temporally localized around $u_*$. Let us distinguish the frequency at which the wave packet is centered, denoting it $\omega$, from the dummy label $\bar \omega$ of generic plane-wave-like modes. Being mathematically more precise, what one has to calculate is actually a wave-packet representation of the Bogoliubov coefficients~\cite{packet}:
\begin{eqnarray}
\beta_{\rm WPR}(u_ *;\omega,\omega') = \int_{-\infty}^{+\infty} d\bar\omega P_{u_ *;\omega}(\bar\omega)\, \beta(u_*;\bar\omega,\omega'). 
\end{eqnarray}
Here $P_{u_*;\omega}(\bar\omega)$ defines the frequency distribution of the wave packets $\phi^{u_*}(\omega;t,\vec x)$,
\begin{eqnarray}
\phi^{u_*}(\omega;t,\vec x) = \int_{-\infty}^{+\infty}  d \bar\omega P_{u_*;\omega}(\bar\omega)\, \phi^u(\bar\omega;t,\vec x).
\end{eqnarray}
These wave packets are taken to be normalized so that
\begin{eqnarray}
\int_{-\infty}^{+\infty}  |P_{u_*;\omega}(\bar\omega)|^2 d\bar\omega = 1.
\end{eqnarray}
In the vicinity of $\omega$ the function $P_{u_*,\omega}(\bar\omega)$ can always be approximated by a slowly varying absolute value multiplied by a rapidly varying phase of the form $e^{i\omega u_*}$. For instance, take a simple wave packet representation
\begin{eqnarray}
P_{u_*;\omega}(\bar\omega)={1 \over \Delta \omega}
\left[\Theta(\bar\omega - \omega +\Delta \omega/2) - \Theta(\bar\omega -\omega-\Delta \omega/2)\right]
e^{i\bar\omega u_*},
\end{eqnarray}
with $\Delta \omega$ the width of the wave packet. 

Now, in all the previous calculations we were neglecting an irrelevant phase factor in $\beta(u_ *;\bar \omega,\omega')$. It is easy to check that this phase factor always contains a term of the form
\begin{eqnarray}
\exp\left( -i{\bar\omega \over \kappa_*} \ln [\omega'  A_*\kappa_*/\bar \omega]  \right).
\end{eqnarray}
Taking this into account, it is not difficult to see that for $\omega \gg \Delta \omega$ an
approximate evaluation (the slowly varying factors that depend on $\bar\omega$ are approximated by their value at $\omega$) 
of $\beta_{\rm WPR}(u_ *;\omega,\omega')$ yields
\begin{eqnarray}
\beta_{\rm WPR}(u_ *;\omega,\omega') 
&\approx& \left|\beta(u_ *;\omega,\omega')\right| \;\;
{\sin \left\{ \Delta\omega \left(u_*- {1\over\kappa_*} \ln [\omega' A_*\kappa_*/\omega] \right)\right\}  \over
\Delta \omega \left( u_*- {1\over\kappa_*} \ln [\omega' A_*\kappa_*/\omega]\right)}
\nonumber\\
&= &
\left|\beta(u_ *;\omega,\omega')\right| \;\;
{\sin\left(- [\Delta\omega/\kappa_*]\; \ln [\omega' C_*/\omega] \right) \over
\left(- [\Delta\omega/\kappa_*]\; \ln [\omega' C_*/\omega]\right)}.
\label{E:WPR}
\end{eqnarray}
Then the number of particles (wave-packets)  of frequency $\omega$ arriving at infinity at time $u_*$ is
\begin{eqnarray}
\int d\omega' |\beta_{\rm WPR}(u_ *;\omega,\omega' )|^2 &\approx&
{1 \over \exp(2\pi\omega/\kappa_*)-1}\;
\nonumber\\
&& \times 
{1 \over 2\pi \kappa_*} \;
\int_{-\infty}^{+\infty} {d\omega' \over \omega'} \; 
{\sin^2\left(-[\Delta\omega/\kappa_*]\;\ln [\omega' C_* /\omega]\right) \over \left(- [\Delta\omega/\kappa_*]\; \ln [\omega' C_* /\omega]\right)^2}
\nonumber \\
&=&
{1 \over \exp(2\pi\omega/\kappa_*)-1},
\end{eqnarray}
which is finite and has precisely the anticipated Planck shape.
 
Furthermore, the function $\sin^2(x)/x^2$ is tightly peaked around $x=0$, with a ``width'' characterized by the location of the first zero at $|x|=\pi$.  So the range of ``important'' values of $\omega'$ contributing to this integral is characterized by 
\begin{equation}
[\Delta\omega/\kappa_*]\; \left| \ln (\omega' C_* /\omega) \right| \lesssim \pi.
\end{equation}
That is
\begin{equation}
e^{-\pi \kappa_*/\Delta\omega} \lesssim {\omega' C_* \over \omega} \lesssim e^{\pi \kappa_*/\Delta\omega}.
\end{equation}
For physically reasonable wave-packets, with physically reasonable widths $\Delta \omega$, this interval is well inside the interval of mathematical validity of the stationary phase calculation as estimated in (\ref{E:range1}), (\ref{E:range2}), so in (\ref{E:WPR}) it is perfectly justified to use a $\beta(u_*;\omega,\omega')$ calculated through this approximation technique. Note that for wave-packets the frequency $\omega'$ on $\scri^-$ is related to the central frequency $\omega$ on $\scri^+$ by the redshift factor $C_*$ and an envelope depending on the frequency width $\Delta \omega$ of the wave-packet.

\section{Physical necessity of the adiabatic condition}

We have now, subject to the approximation underlying the exponential approximation, which we emphasize is always valid on a small enough region, derived the existence of a (slowly varying) flux of particles arriving at $\scri^+$ with a characteristic temperature  $k_B \;T_H(u)=  \hbar \; \kappa(u)/2\pi$. The derivation was ultimately based on wave-packets of support confined to the region  ${\cal S}_+$ over which the exponential approximation is valid. 
But a wave-packet supported on $ {\cal S}_+$ is bandwidth limited to only contain a limited range of frequencies
\begin{equation}
\Delta \omega \gtrsim { \sqrt{|\dot\kappa_*|}\over \epsilon}.
\end{equation}
If we wish the wave-packet to contain frequencies capable of probing the  Planck peak of the Hawking spectrum (instead of merely probing the high-frequency Boltzmann tail) then we must have $\Delta\omega\lesssim\kappa_*$.  Then:
\begin{equation}
{\sqrt{|\dot\kappa_*|}\over \epsilon} \lesssim \kappa_*; \qquad {|\dot\kappa_*|\over\kappa_*^2} \lesssim \epsilon^2; \qquad 
 {|\dot\kappa_*|\over\kappa_*^2} \ll 1.
\end{equation}
So here we have another viewpoint on the necessity of the adiabatic condition: While by direct calculation the adiabatic condition is seen to be automatically valid for macroscopic (Schwarzschild) black holes, the reason the condition is useful is that it permits the calculation to probe the Planck peak of the spectrum.

\section{Discussion}

Note what we think is the importance of these results ---  the formalism we have developed is an extremely general way of getting
Hawking-like fluxes from a rather broad class of spacetimes --- of
course it agrees with previous results and  also puts things into a more general
framework. The key result is this:

Whenever $\scri^-$ and $\scri^+$ are connected by a function $U=p(u)$ of the form
\begin{equation}
U = U_* 
+ C_* \int^u_{u^*} \;   
\exp\left[  - \int^{\bar u}_{u_*} \kappa(\tilde u) \; d\tilde u \right] d  \bar u,
\end{equation}
for some constant $C_*$, and where $\kappa(u)$ satisfies an ``adiabatic
condition''
\begin{equation}
\left|{d\kappa(u)\over du}\right|_{u_*}  \ll \kappa(u_*)^2,
\end{equation}
then the argument sketched above proves the existence of a
Hawking-like Planck-distributed spectrum of outgoing particles reaching $\scri^+$ at
$u_*$ with a (time-dependent) Hawking temperature
\begin{equation}
k_B \; T_H( u_* ) = {\hbar \; \kappa( u_* )\over2\pi}.
\end{equation}
As we move along $\scri^+$ (that is, as $u_*$ increases, possibly even
with $u_* \to +\infty$), this relation continues to hold, with the
Hawking temperature controlled by $\kappa(u_*)$,  \emph{as long as the
adiabatic condition continues to hold}.

Working with this formalism we have never had to use any statements about any horizon forming anywhere --- neither event horizon nor apparent horizon nor trapping horizon nor dynamical horizon --- so this is consistent with all relevant previously known results~\cite{hawking1, hawking2, unruh, tipler, israel, bardeen, york, hajicek,  grove, entropy, ergo,  essential, lindesay+sheldon,  dbh,  bergmann-roman,  quasiparticle, trapping,  fate,  small-dark, revisit, hawking-dublin, hawking-post-dublin, ashtekar-bojowald, disinformation, hayward,  our-prl-article}. For example, within this formalism one can check that the standard picture of Hawking evaporation is fully justified, even without making any particular commitment (pro or con) as to event-horizon formation.
As the black hole loses mass its ``surface gravity'' $\kappa$ rises, and so does its Hawking temperature. And this picture should continue to hold all the way to
the Planck scale. (At which stage it is the adiabatic condition that breaks down, along with the related breakdown of validity of the exponential approximation --- ``merely'' indicating the need for a more thorough fully dynamic analysis.)
The current formalism in no way ``solves'' the trans-Planckian problem and if anything brings it more sharply into focus. Over the years there have been repeated suggestions that Hawking emission should in some sense be ``localized'' (or more precisely ``delocalized'') to within an $e$-folding distance or two of the horizon (or in our case the might-be horizon).  For the current state of affairs see~\cite{localize, localize2}.  Note that we have carefully described the Hawking flux as Planckian rather than thermal. To claim thermality one has to explicitly assume the formation of an event horizon (behind which one can hide correlations). 

Finally, it is interesting to remark that within this formalism other non-standard evaporation scenarios can also be envisaged and analyzed. For example, the evaporation of hypothetical compact horizonless objects follows the described pattern, with a $\kappa(u)$ differing at each stage from $\kappa_{\rm Hawking}$~\cite{quasiparticle, trapping}. In addition, there is a chance that in these scenarios the evaporation process can proceed all the way to its end-point in a completely adiabatic manner~\cite{quasiparticle, trapping}. Work on these important points is ongoing.

\clearpage
\appendix
\section{Surface gravity: peeling versus inaffinity}
\def\d{d}

In general, the peeling notion of surface gravity $\kappa_\mathrm{peeling}$ is defined by the ``peeling off'' properties of null geodesics near the horizon, while the more usual textbook notion of surface gravity $\kappa_\mathrm{inafinity}$ is defined by the ``inaffinity'' of the naturally normalised null geodesic at the horizon. For Killing horizons these concepts coincide.
For dynamical/ trapping/ apparent/ evolving/ putative horizons they will in general differ. In this appendix we shall motivate the relevant definitions --- we will avoid the use of null coordinates, and adopt the more usual Schwarzschild curvature coordinates. 

For the static case, write the metric in the form~\cite{dbh}
\begin{equation}
\d s^2 = - e^{-2\Phi(r)} [1-2m(r)/r)] \d t^2 + {\d r^2\over1-2m(r)/r} + r^2 \{ \d\theta^2+\sin^2\theta\; \d\phi^2\}.
\end{equation}
The Killing horizon is defined by the location where $2m(r)/r=1$, that is
\begin{equation}
2m(r_H)=r_H.
\end{equation}
Then it is an old result~\cite{dbh} that
\begin{equation}
\kappa = {e^{-\Phi_H} (1-2m'_H)\over 2 r_H}.
\end{equation}

\subsection{The peeling notion of surface gravity}

For the evolving case, write the metric in the form~\cite{alex, abreu, abreu2} 
\begin{equation}
\d s^2 = - e^{-2\Phi(r,t)} [1-2m(r,t)/r)] \d t^2 + {\d r^2\over1-2m(r,t)/r} + r^2 \{ \d\theta^2+\sin^2\theta\; \d\phi^2\},
\end{equation}
and define the ``evolving horizon'' by the location where $2m(r,t)/r = 1$, that is
\begin{equation}
2 m(r_H(t),t) = r_H(t).
\end{equation}
To calculate $\kappa_\mathrm{peeling}$ note that a radial null geodesic satisfies
\begin{equation}
\left({\d r\over\d t}\right)^2 =  e^{-2\Phi(r,t)} [1-2m(r,t)/r)]^2,
\end{equation}
that is
\begin{equation}
\left({\d r\over\d t}\right) =  \pm e^{-\Phi(r,t)} [1-2m(r,t)/r)].
\end{equation}
Now if the geodesic is near $r_H(t)$, that is  $r \approx r_H(t)$, we can Taylor expand
 \begin{equation}
{\d r\over\d t} =  \pm {e^{-\Phi(r_H(t),t)} [1-2m'(r_H(t),t)]\over r_H(t)} \; [r(t) - r_H(t)] +{\cal O}\left( [r(t) - r_H(t)]^2 \right),
\end{equation}
where now a prime denotes derivative with respect to $r$.
That is, defining
\begin{equation}
\kappa_\mathrm{peeling}(t) =  {e^{-\Phi(r_H(t),t)} [1-2m'(r_H(t),t)]\over 2 r_H(t)},
\end{equation}
we have
 \begin{equation}
{\d r\over\d t} =  \pm 2 \kappa_\mathrm{peeling}(t) \; [r(t) - r_H(t)] +{\cal O}\left( [r(t) - r_H(t)]^2 \right).
\end{equation}
Then for two null geodesics $r_1(t)$ and $r_2(t)$ close to and on the same side of the evolving horizon
\begin{equation}
{\d (r_1-r_2)\over\d t} \approx  \pm  2\kappa_\mathrm{peeling}(t) \; [r_1(t) - r_2(t)],
\end{equation}
which it is better to write as
\begin{equation}
{\d |r_1-r_2|\over\d t} \approx 2 \kappa_\mathrm{peeling}(t) \; |r_1(t) - r_2(t)|,
\end{equation}
because this automatically keeps track of all the signs. 
Therefore
\begin{equation}
|r_1(t)-r_2(t)| \approx |r_1(t_0)-r_2(t_0)| \; \exp \left[ 2\int \kappa_\mathrm{peeling}(t) \; \d t \right].
\end{equation}
So $\kappa_\mathrm{peeling}(t)$ is a natural time-dependent generalization of the static version of $ \kappa_\mathrm{peeling}$ which correctly encodes the exponential ``peeling off'' behaviour of null geodesics. It is this quantity that we have seen is ultimately connected to the temperature of the Hawking flux.

\subsection{The inaffinity notion of surface gravity}
\label{A:2}

In contrast, to calculate $\kappa_\mathrm{inaffinity}$  consider the outward-pointing radial null vector field
\begin{equation}
\ell^a = \left( 1,  e^{-\Phi(r,t)} (1-2m(r,t)/r), 0, 0\right).
\end{equation}
In a static spacetime, this null vector field is simply related to the Killing vector $K^a$ as
\begin{equation}
\ell^a = K^a + \epsilon^a{}_b K^b,
\end{equation}
where $\epsilon^a{}_b$ is the 2-dimensional Levi--Civita tensor on the $\{r,t\}$ plane.
Hence the inaffinity  $\kappa_\mathrm{bulk}(r,t)$ defined by 
\begin{equation}
\ell^a \nabla_a \ell^b = 2\, \kappa_\mathrm{bulk}(r,t) \; \ell^b,
\end{equation}
which \emph{always} exists, 
everywhere throughout the spacetime, now naturally defines a notion of generalized surface gravity even for a time-dependent geometry.

A brief calculation~\cite{alex, abreu, abreu2} shows that
\begin{equation}
\kappa_\mathrm{bulk}(r,t) = {e^{-\Phi} [m(r,t)/r -  m'(r,t)] \over r} -  e^{-\Phi} \Phi'(r,t) [1-2m(r,t)/r] - {1\over2}\dot\Phi(r,t).
\end{equation}
Note that at the evolving horizon
\begin{eqnarray}
\kappa_\mathrm{bulk}(r_H(t),t) &=& {e^{-\Phi(r_H(t),t)} [1 - 2 m'(r_H(t),t)] \over 2r}  - {1\over2}\dot\Phi(r_H(t),t)  
\nonumber\\
&=& \kappa_\mathrm{peeling}(t)  - {1\over2}\dot\Phi(r_H(t),t).  
\end{eqnarray}
But this is not yet $\kappa_\mathrm{inaffinity}(t)$. 
The quantity $\kappa_\mathrm{inaffinity}(t)$ is found by somehow locating the true event horizon $r_E(t) \neq r_H(t)$ (assuming an event horizon forms) and defining
\begin{equation}
\kappa_\mathrm{inaffinity}(t) = \kappa_\mathrm{bulk}(r_E(t),t).
\end{equation}
While we do not know exactly where the event horizon is, we can hope that when asymptotically approaching a  quasi-static situation we have
\begin{equation}
r_E(t) \approx r_H(t),
\end{equation}
in which case we can expand in a Taylor series
\begin{equation}
\kappa_\mathrm{inaffinity}(t)\approx  \kappa_\mathrm{bulk}(r_H(t),t) +  \kappa_\mathrm{bulk}'(r_H(t),t) [r_E(t)-r_H(t)].
\end{equation}
That is
\begin{equation}
\kappa_\mathrm{inaffinity}(t)\approx   \kappa_\mathrm{peeling}(t)  - {1\over2}\dot\Phi(r_H(t),t)   +  \kappa_\mathrm{bulk}'(r_H(t),t) [r_E(t)-r_H(t)].
\end{equation}
This clearly demonstrates that the two concepts are in general distinct, though they will automatically agree in static (or stationary) spacetimes.
Calculating the quantity $\kappa_\mathrm{bulk}'(r_H(t),t)$ is do-able but unedifying.

\section*{Acknowledgements}

The research of Matt Visser was supported by the Marsden Fund administered by the Royal Society of New Zealand. MV also wishes to thank the \emph{Instituto de Astrof\'\i{}sica de Andaluc\'ia} (IAA--CSIC) and the \emph{Scuola Internazionale Superiore di Studi Avanzati} (SISSA) for hospitality during various phases of this project, and \emph{INFN (Sezione di Trieste)} for partial financial support. 

Carlos Barcel\'o has been supported by the Spanish MICINN through the project FIS2008-06078-C03-01 and by the Junta de Andaluc\'{\i}a through the projects FQM2288 and FQM219.

Stefano Liberati and Sebastiano Sonego   wish to thank Stefano Finazzi for useful insights and support in testing some of the  ideas presented herein. SL also wishes to thank Renaud Parentani for illuminating remarks concerning  various notions of surface gravity and their historical evolution.



\begin{thebibliography}{666}
\enlargethispage{20pt}
\vskip-15pt
\bibitem{hawking1}%
S.~W.~Hawking, %
``Black hole explosions'', %
Nature {\bf 248} (1974) 30--31.%
\bibitem{hawking2}%
S.~W.~Hawking, %
``Particle creation by black holes'', %
Commun.\ Math.\ Phys.\ {\bf 43} (1975) 199--220; %
Erratum: {\em ibid.\/} {\bf 46} (1976)206. %
\bibitem{unruh}
W.~G.~Unruh,
  ``Notes on black hole evaporation'',
  Phys.\ Rev.\ D {\bf 14} (1976) 870.
\bibitem{tipler}%
F.~J.~Tipler, %
``Do black holes really evaporate thermally?'' %
Phys.\ Rev.\ Lett.\ {\bf 45} (1980) 949--951.%
\bibitem{israel}%
P.~Hajicek and W.~Israel, %
``What, no black hole evaporation?'' %
Phys.\ Lett.\ A {\bf 80} (1980) 9--10.  %
\bibitem{bardeen}%
J.~M.~Bardeen, %
``Black holes do evaporate thermally'', %
Phys.\ Rev.\ Lett.\ {\bf 46} (1981) 382--385.%
\bibitem{york}%
J.~W.~York, Jr., %
``Dynamical origin of black-hole radiance'', %
Phys.\ Rev.\ D {\bf 28} (1983) 2929--2945.%
\bibitem{hajicek}
P.~Hajicek, 
 ``Origin of Hawking radiation''
 Phys.\ Rev.\ D {\bf 36} (1987) 1065--1079.
\bibitem{grove}%
P.~G.~Grove, %
``Observations on particle creation by static gravitational fields'', %
Class.\ Quantum Grav.\ {\bf 7} (1990) 1353--1363.%
\bibitem{entropy}
M.~Visser,
  ``Hawking radiation without black hole entropy'',
  Phys.\ Rev.\ Lett.\  {\bf 80} (1998) 3436
  [arXiv:gr-qc/9712016].
\bibitem{ergo}
M.~Visser,
  ``Acoustic black holes: Horizons, ergospheres, and Hawking radiation'',
  Class.\ Quant.\ Grav.\  {\bf 15} (1998) 1767
  [arXiv:gr-qc/9712010].
\bibitem{essential}
M.~Visser,
  ``Essential and inessential features of Hawking radiation'',
  Int.\ J.\ Mod.\ Phys.\ D {\bf 12} (2003) 649
  [arXiv:hep-th/0106111].
\bibitem{lindesay+sheldon}
  J.~Lindesay and P.~Sheldon,
  ``Penrose diagram for a transient black hole'',
  Class.\ Quant.\ Grav.\  {\bf 27 } (2010)  215015.
  [arXiv:1005.4449 [gr-qc]].
\bibitem{Parikh:1999}
  M.~K.~Parikh and F.~Wilczek,
  ``Hawking radiation as tunneling'',
  Phys.\ Rev.\ Lett.\  {\bf 85} (2000) 5042
  [arXiv:hep-th/9907001].
  \bibitem{Parikh:2004a}
  M.~K.~Parikh,
  ``Energy conservation and Hawking radiation'',
  [hep-th/0402166].
\bibitem{Parikh:2004b}
  M.~K.~Parikh,
  ``A Secret tunnel through the horizon'',
  Int.\ J.\ Mod.\ Phys.\  {\bf D13 } (2004)  2351-2354.
  [hep-th/0405160].  
 \bibitem{Shankaranarayanan:2000}
  S.~Shankaranarayanan, T.~Padmanabhan, K.~Srinivasan,
  ``Hawking radiation in different coordinate settings: Complex paths approach'',
  Class.\ Quant.\ Grav.\  {\bf 19}, 2671-2688 (2002).
  [gr-qc/0010042].
 \bibitem{Angheben:2005}
  M.~Angheben, M.~Nadalini, L.~Vanzo, and S. Zerbini,
  ``Hawking radiation as tunneling for extremal and rotating black holes'',
  JHEP {\bf 0505 } (2005)  014.
  [hep-th/0503081].
  \bibitem{Medved:2005}
  A.~J.~M.~Medved, E.~C.~Vagenas,
  ``On Hawking radiation as tunneling with back-reaction'',
  Mod.\ Phys.\ Lett.\  {\bf A20 } (2005)  2449-2454.
  [gr-qc/0504113].
  \bibitem{Arzano:2005}
  M.~Arzano, A.~J.~M.~Medved, E.~C.~Vagenas,
  ``Hawking radiation as tunneling through the quantum horizon'',
  JHEP {\bf 0509 } (2005)  037.
  [hep-th/0505266].
  \bibitem{Robinson:2005}
  S.~P.~Robinson, F.~Wilczek,
  ``A Relationship between Hawking radiation and gravitational anomalies'',
  Phys.\ Rev.\ Lett.\  {\bf 95 } (2005)  011303.
  [gr-qc/0502074].
 \bibitem{Clifton:2008}
 T.~Clifton,
 ``Properties of Black Hole Radiation From Tunnelling'',
 Class.\ Quant.\ Grav.\  {\bf 25}, 175022 (2008).
 [arXiv:0804.2635 [gr-qc]].
\bibitem{Banerjee:2009}
R.~Banerjee, B.~R.~Majhi,
``Hawking black body spectrum from tunneling mechanism'',
 Phys.\ Lett.\  {\bf B675}, 243-245 (2009).
 [arXiv:0903.0250 [hep-th]].
 \bibitem{Padmanabhan:2003}
  T.~Padmanabhan,
  ``Gravity and the thermodynamics of horizons'',
  Phys.\ Rept.\  {\bf 406 } (2005)  49-125.
  [gr-qc/0311036].
\bibitem{Hossenfelder:2002}
  S.~Hossenfelder, D.~J.~Schwarz and W.~Greiner,
  ``Particle production in time-dependent gravitational fields: The expanding mass shell'',
  Class.\ Quant.\ Grav.\  {\bf 20} (2003) 2337
  [arXiv:gr-qc/0210110].
\bibitem{dbh}
M.~Visser,
  ``Dirty black holes: Thermodynamics and horizon structure'',
  Phys.\ Rev.\ D {\bf 46} (1992) 2445
  [arXiv:hep-th/9203057].
\bibitem{bergmann-roman}
T.~A.~Roman and P.~G.~Bergmann,  
``Stellar collapse without singularities?'', 
Phys. Rev. {\bf D28} (1983)  1265--1277.
  \bibitem{quasiparticle} 
 C.~Barcel\'o, S.~Liberati, S.~Sonego, and M.~Visser,
  ``Quasi-particle creation by analogue black holes'',
  Class.\ Quant.\ Grav.\  {\bf 23} (2006) 5341
  [arXiv:gr-qc/0604058].
  \bibitem{trapping}
  C.~Barcel\'o, S.~Liberati, S.~Sonego, and M.~Visser,
  ``Hawking-like radiation does not require a trapped region'',
  Phys.\ Rev.\ Lett.\  {\bf 97} (2006) 171301
  [arXiv:gr-qc/0607008].
 \bibitem{fate} 
C.~Barcel\'o, S.~Liberati, S.~Sonego, and M.~Visser,
``Fate of gravitational collapse in semiclassical gravity'', 
Phys.\ Rev.\ D {\bf 77} (2008) 044032
[arXiv:0712.1130 [gr-qc]].
\bibitem{small-dark} 
M.~Visser, C.~\Barcelo, S.~Liberati, and S.~Sonego,
`Small, dark, and heavy: But is it a black hole?'',
PoS (BHs GR \& Strings) 2009, 010.
  [arXiv:0902.0346 [gr-qc]].
\bibitem{revisit}
C.~\Barcelo, S.~Liberati, S.~Sonego, and M.~Visser,
 ``Revisiting the semiclassical gravity scenario for gravitational collapse'',
  AIP Conf.\ Proc.\  {\bf 1122} (2009) 99-106.
  [arXiv:0909.4157 [gr-qc]].
\bibitem{hawking-dublin}
S. W. Hawking, abstract of talk at GR17, Dublin, Ireland, July 2004: 
``The way the information gets out seems to be that a true event horizon never forms, just an 
apparent horizon.''
\bibitem{hawking-post-dublin}
S. W. Hawking, ÒInformation loss in black holesÓ, 
Phys. Rev.  D {\bf 72} (2005) 084013 [arXiv:hep-th/0507171]. 
 \bibitem{ashtekar-bojowald}
  A.~Ashtekar and M.~Bojowald,
  ``Black hole evaporation: A paradigm'',
  Class.\ Quant.\ Grav.\  {\bf 22} (2005) 3349
  [arXiv:gr-qc/0504029].
  \bibitem{disinformation}
  S.~A.~Hayward,
  ``The disinformation problem for black holes'',
  arXiv:gr-qc/0504037; gr-qc/0504038
  \bibitem{hayward}
  S.~A.~Hayward,
  ``Formation and evaporation of regular black holes'',
  Phys.\ Rev.\ Lett.\  {\bf 96} (2006) 031103
  [arXiv:gr-qc/0506126].
\bibitem{unexpected}
  M.~Visser,
  ``Acoustic propagation in fluids: An unexpected example of Lorentzian geometry'',
  [gr-qc/9311028].
 \bibitem{LRR}
  C.~\Barcelo, S.~Liberati, and M.~Visser,
  ``Analogue gravity'',
  Living Rev.\ Rel.\  {\bf 8 } (2005)  12
  [gr-qc/0505065].
\bibitem{our-prl-article}
C.~\Barcelo, S.~Liberati, S.~Sonego, and M.~Visser,
``Minimal conditions for the existence of a Hawking-like flux'',
[arXiv:1011.5593 [gr-qc]].
\bibitem{hu}%
B.~L.~Hu, %
``Hawking-Unruh thermal radiance as relativistic exponential scaling of quantum noise'', %
in {\em Thermal Field Theory and Applications\/}, %
edited by Y.\ X.\ Gui, F.\ C.\ Khanna, and Z.\ B.\ Su %
(Singapore, World Scientific, 1996), pp.\ 249--260.
[arXiv:gr-qc/9606073].%
\bibitem{njp}
C.~\Barcelo, S.~Liberati, S.~Sonego, and M.~Visser,
  ``Causal structure of acoustic spacetimes'',
  New J.\ Phys.\  {\bf 6 } (2004)  186.
  [gr-qc/0408022].
\bibitem{Macher}
  J.~Macher and R.~Parentani,
  ``Black-hole radiation in Bose-Einstein condensates'',
  Phys. Rev. {\bf A80} (2009)  043601.
  [arXiv:0905.3634 [cond-mat.quant-gas]].
\bibitem{Brout1}
R.~Brout, S.~Massar, R.~Parentani and P.~Spindel,
  ``Hawking Radiation Without Transplanckian Frequencies'',
  Phys.\ Rev.\  D {\bf 52} (1995) 4559
  [arXiv:hep-th/9506121].
\bibitem{Brout2}
  R.~Brout, S.~Massar, R.~Parentani and Ph.~Spindel,
  ``A Primer for Black Hole Quantum Physics'',
  Phys.\ Rept.\  {\bf 260} (1995) 329
  [arXiv:0710.4345 [gr-qc]].
  \bibitem{barrow}
  J.~D.~Barrow,
  ``Sudden future singularities'',
  Class.\ Quant.\ Grav.\  {\bf 21 } (2004)  L79-L82.
  [gr-qc/0403084].
  \bibitem{barrow2}
  J.~D.~Barrow,
  ``More general sudden singularities'',
  Class.\ Quant.\ Grav.\  {\bf 21 } (2004)  5619-5622.
  [gr-qc/0409062].
\bibitem{cattoen}
  C.~Catto\"en and M.~Visser,
  ``Necessary and sufficient conditions for big bangs, bounces, crunches, rips, sudden singularities, and extremality events'',
  Class.\ Quant.\ Grav.\  {\bf 22 } (2005)  4913-4930.
  [gr-qc/0508045].
\bibitem{thunderbolt}
C.~G.~Callan, Jr., S.~B.~Giddings, J.~A.~Harvey, and A.~E.~Strominger,
 ``Evanescent black holes'',
 Phys.\ Rev.\  {\bf D45 } (1992)  1005-1009.
 [hep-th/9111056].
\bibitem{thunderbolt2}
 S.~W.~Hawking and J.~M.~Stewart,
 ``Naked and thunderbolt singularities in black hole evaporation'',
 Nucl.\ Phys.\  {\bf B400 } (1993)  393-415.
 [hep-th/9207105].
\bibitem{birrell-davies}
N.~D.~Birrell and P.~C.~W.~Davies, %
{\em Quantum Fields in Curved Space\/} %
(Cambridge University Press, 1982).%
\bibitem{packet}
C.~\Barcelo, L.~J.~Garay, and G.~Jannes,
  ``Sensitivity of Hawking radiation to superluminal dispersion relations'',
  Phys.\ Rev.\  {\bf D79} (2009) 024016.
  [arXiv:0807.4147 [gr-qc]].
 \bibitem{localize}
R.~Sch\"utzhold and W.~G.~Unruh,
  ``On the origin of the particles in black hole evaporation'',
  Phys.\ Rev.\  D {\bf 78} (2008) 041504
  [arXiv:0804.1686 [gr-qc]].
 \bibitem{localize2}
 W.~G.~Unruh,
  ``Where are the particles created in black hole evaporation'',  
  PoS (QG-Ph) 039.
\bibitem{alex}
A.~B.~Nielsen and M.~Visser,
  ``Production and decay of evolving horizons'',
  Class.\ Quant.\ Grav.\  {\bf 23} (2006) 4637
  [arXiv:gr-qc/0510083].
\bibitem{abreu}
G.~Abreu and M.~Visser,
  ``Kodama time: Geometrically preferred foliations of spherically symmetric spacetimes'',
  Phys.\ Rev.\  D {\bf 82} (2010) 044027
  [arXiv:1004.1456 [gr-qc]].
\bibitem{abreu2}
G.~Abreu and M.~Visser,
  ``Tolman mass, generalized surface gravity, and entropy bounds'',
  Phys.\ Rev.\ Lett.\  {\bf 105} (2010) 041302
  [arXiv:1005.1132 [gr-qc]].

\end{thebibliography}
\end{document}